\documentclass[preprint,nofootinbib,aps,superscriptaddress]{revtex4}
\usepackage{epsfig}
\usepackage{graphicx}
\usepackage{amsmath}
\usepackage{feynmf}
\usepackage{changes}
\def\be{\begin{equation}} 
\def\ee{\end{equation}} 
\def\bea{\begin{eqnarray}}
\def\eea{\end{eqnarray}} 
\def\nnb{\nonumber}

\begin{document}
\title{
Neutrino-Deuteron Reactions at Solar Neutrino Energies 
in Pionless Effective Field Theory with Dibaryon Fields
}

\author{Shung-Ichi Ando}
\email{sando@sunmoon.ac.kr}
\affiliation{School of Mechanical and ICT Convergence Engineering, 
Sunmoon University, Asan, Chungnam 31460, Korea}
\author{Young-Ho Song}
\email{yhsong@ibs.re.kr}
\affiliation{Rare Isotope Science Project, 
Institute for Basic Science, Daejeon 34047, Korea}
\author{Chang Ho Hyun}
\email{hch@daegu.ac.kr}
\affiliation{Department of Physics Education, Daegu University,
Gyeongsan 38453, Korea}

\date{\today}

\begin{abstract}

We study breakup of the deuteron induced by neutrinos in the neutral 
$\nu d\to \nu np$, $\bar{\nu} d\to \bar{\nu} np$ and the charged
$\bar{\nu} d\to e^+ n n$, $\nu d\to e^- pp$ processes.
Pionless effective field theory with dibaryon fields 
is used to calculate the total cross
sections for neutrino energies $E_\nu$ from threshold to 20 MeV.
Amplitudes are expanded up to next-to-leading order, and the partial wave
is truncated at $P$-waves.
The Coulomb interaction between two protons is 
included nonperturbatively in the reaction amplitudes,
and an analytic expression of the amplitudes 
is obtained.  
The contribution of the next-to-leading order to the total cross section 
is in the range of 5.2$-$9.9\% in magnitude, 
and that of the $P$-wave is 2.4$-$2.8\% at $E_\nu = 20$ MeV.  
Uncertainty arising from an axial isovector low-energy constant is estimated to be on 
the order of 1\%.

\end{abstract}
%

\maketitle

\section{Introduction} 

Since first postulated by W. Pauli, neutrinos have played key roles 
in understanding the nature of weak interactions 
and in testing the standard model~\cite{dgh2014}.
These fundamental questions can be answered directly 
or indirectly by measuring the cross sections of neutrinos.
Measurements have been conducted with neutrinos from artificial sources 
and various events in the Universe.
The corresponding energy scale is indeed wide, ranging 
from eV to EeV~\footnote{
EeV is an initialism of exa-electron volt, $10^{18}$~eV.}.
Neutrinos in the energy range from 1 to 20 MeV are particularly important 
in probing the solar process and flavor oscillation
of neutrinos.
Solar neutrinos 
on a deuteron target were measured at Sudbury Neutrino
Observatory (SNO),
and the neutrino flavor mixing parameters were deduced 
by using the $\nu d$ cross sections estimated 
by theory~\cite{sno1,sno2,sno3,sno4}.
Meanwhile,
%
direct measurements on deuteron targets 
in the solar neutrino range have been carried out 
in reactor experiments.
Cross sections were reported for the charged-current process 
$\bar{\nu}_e d \to e^+ nn$, and the neutral-current process 
$\bar{\nu}_e d \to \bar{\nu}_e n p$~\cite{pas1979,ver1991,koz2000,ril1999}.
Comparison of experiments with theory exhibits qualitative 
agreement~\cite{for2012}.

Early theoretical estimates of neutrino reactions on the deuteron 
were reported in the 1960s~\cite{ku-prl66,eb-npa68} 
while modern ones are improved
by using accurate phenomenological nucleon-nucleon ($NN$) potentials
and including meson exchange currents~\cite{tkk-prc90,snpa,netal-npa}.
Modern theories of 
$NN$ interactions can reproduce the $NN$ scattering phase shift data
below the pion production threshold with errors less than 1\%.
These high-precision theories provide a unique opportunity to probe the interactions of neutrino 
and deuteron with uncertainties due to strong interactions under control.
In publications during the last decade, we have applied the pionless
effective field theory with dibaryon fields 
(dEFT in short)~\cite{k-npb97,bs-npa01}
 to low-energy two-nucleon systems 
and phenomena such as electromagnetic (EM) form factors of the
deuteron \cite{prc2005}, synthesis of the deuteron at big bang energies \cite{prc2006},
proton-proton scattering \cite{prc2007}, their fusion \cite{plb2008}, neutron-proton scattering \cite{prc2012}, 
spin-dependent polarization \cite{prc2011,fb2013spin, prc2017}, and hadronic parity violation 
in radiative neutron-proton fusion or
the dissociation of the deuteron \cite{mpla2009, prc2010, npa2010,fb2013pv, prc2013}.
We verified that (i) calculational complexity and difficulty are significantly reduced in dEFT
compared to the calculations in phenomenological potential models or other EFTs,
(ii) convergence of the expansion is fast so only up to next-to-next-to-leading order (NNLO)
results agree with high-quality calculation of phenomenological models, and
(iii) agreement with experiment and other theories is achieved with difference less than 1\% at low energies.

Inspired by successful applications to the strong and EM interactions and related phenomena,
here we attempt to apply the model to a semi-leptonic weak interaction problem, breakup of the
deuteron by neutrinos at low energies. 
The solution of the problem plays an important role in understanding the flavor oscillation of neutrinos 
from the sun and in testing the validity of the standard model.
In addition, it has been discussed that the process can have a non-negligible effect on the supernova
neutrino emission mechanism \cite{nasu}.
The problem has been explored in diverse frameworks, 
such as a conventional approach using the two-nucleon potential
plus meson exchange currents \cite{snpa}, hybrid EFT \cite{hybrid},
pionless EFT \cite{butler}, and a recent work with chiral perturbation theory \cite{baroni}.
The results in various theoretical works agree to high accuracy.

In this work we focus on two issues: 
(i) estimating the uncertainty of the prediction and 
(ii) investigating the convergence and accuracy of the expansion in the dEFT formalism.
We calculate the total cross sections of the neutral-current (NC) processes 
\begin{eqnarray}
\nu_l d \to \nu_l np, \, \, 
\bar{\nu}_l d \to \bar{\nu}_l np,
\end{eqnarray} 
where $l$ denotes the lepton flavor $e$, $\mu$ and $\tau$, 
and the charged-current (CC) processes 
\begin{eqnarray}
\nu_e d \to e^- pp, \, \, 
\bar{\nu}_e d \to e^+ nn,
\end{eqnarray}
with neutrino energy from threshold to 20~MeV.
We consider the expansion up to next-to-leading order (NLO) 
and assume a $P$-wave approximation
for the partial-wave expansion in the final state of the nucleon. 
In addition, we include the Coulomb interaction between two protons 
non-perturbatively and obtain an analytic expression for the reaction
amplitudes for the $\nu_e d \to e^- pp$ process.
Results are compared to the most updated calculations with modern potential models and EFTs.

Section II summarizes the basic equations 
and the analytic forms of transition amplitudes and cross sections. 
Numerical results are presented in Sec. III, 
and conclusions are drawn in Sec. IV.
In Appendix A a derivation of an analytic expression of the 
reaction amplitudes for the $\nu_e d\to e^-pp$ process is presented,
and in Appendix B the spin summation relations and an expression of 
the squared amplitudes are displayed. 
In Appendix C, we address a way to determine an axial-isovector 
low-energy constant (LEC) directly from the $\beta$ decay of tritium, 
and we discuss differences from this work.

\section{Analytic Expression for the reaction amplitudes}
\subsection{Weak process and dibaryon EFT}

The reaction amplitudes 
can be calculated from the effective Hamiltonian 
of the current-current interaction~\cite{dgh2014}
\begin{eqnarray}
H = \frac{G'_F}{\sqrt{2}} \int d^3 x \left[
V_{ud} J^{(CC)}_\mu (\vec{x}) l^{(CC)\mu} (\vec{x}) 
+ J^{(NC)}_\mu l^{(NC) \mu}(\vec{x}) \right],
\end{eqnarray}
where $G'_F = 1.1803 \times 10^{-5}$ GeV$^{-2}$ is the weak coupling  constant,
and $V_{ud} = 0.9746$ is a Cabbibo-Kobayashi-Maskawa matrix element. 
Note that $G'_F$ includes the inner radiative corrections: 
$G'^2_F = G^2_F (1+\Delta^V_R)$, where
$G_F= 1.1166 \times 10^{-5}$ GeV$^{-2}$ 
is the Fermi constant and $\Delta^V_R$ is the
inner radiative correction~\cite{msprl1993,aetalplb2004}.
The CC- and NC-lepton currents $l^{(CC)}_\mu$ and $l^{(NC)}_\mu$ 
are well known, and the CC- and NC-hadron currents $J^{(CC)}_\mu$ 
and $J^{(NC)}_\mu$ are written as
\begin{eqnarray}
J^{(CC)}_\mu &=& V^\pm_\mu (\vec{x}) - A^\pm_\mu(\vec{x}), \\
J^{(NC)}_\mu &=& (1-\sin^2\theta_W) V^0_\mu(\vec{x}) - A^0_\mu(\vec{x}) -
2 \sin^2\theta_W V^S_\mu(\vec{x}),
\end{eqnarray}
where $V_\mu$ and $A_\mu$ represent the vector and axial current, respectively.
The superscripts $\pm$ and 0 are the isospin indices of the isovector current 
and $S$ denotes the isoscalar current. 
$\theta_W$ is the Weinberg angle with the numerical value 
$\sin^2\theta_W=0.2312$.

The CC- and NC-hadronic currents $J_\mu^{(CC)}$ and $J_\nu^{(NC)}$ 
are calculated  from the Lagrangians for the $\nu d$ scattering 
relevant up to NLO~\cite{prc2005,plb2008,nnfusion}
\begin{eqnarray}
{\cal L} = 
{\cal L}_0 
+ {\cal L}_1 
+ {\cal L}_t 
+ {\cal L}_s 
+ {\cal L}_{\rm int},
\label{eq:lagrangian}
\end{eqnarray}
where ${\cal L}_0$ and ${\cal L}_1$ are 
leading order (LO) and NLO one-nucleon Lagrangians,
${\cal L}_s$ and ${\cal L}_t$ are the Lagrangians 
for dibaryon-dibaryon and dibaryon-nucleon couplings
in the $^{1}{\rm S}_0$ and $^{3}{\rm S}_1$ states, respectively,
and ${\cal L}_{\rm int}$ denotes the EM and weak interactions 
of nucleons and dibaryons through external vector and axial-vector 
fields.

The Lagrangian for the one-nucleon sector is given as~\cite{aetalplb2004}
\begin{eqnarray}
{\cal L}_0 &=& N^\dagger \left[ i v \cdot \tilde{D} + 2 g_A S \cdot a \right] N, 
\\
{\cal L}_1 &=& \frac{1}{2 m_N} N^\dagger \left[ (v \cdot \tilde{D})^2 
- \tilde{D}^2 - 2 i g_A \left\{ v \cdot a, S \cdot \tilde{D} \right\} \right.
\nonumber \\
&&   \left. -2 i (1+\kappa_V ) [S^\mu, S^\nu] f^+_{\mu \nu} 
- 2 i (1+\kappa_S) [S^\mu, S^\nu] v_{S \mu \nu} \right]N,
\end{eqnarray}
with
\begin{eqnarray}
\tilde{D}^\mu &=& \partial^\mu - i \frac{1}{2} {\vec\tau}\cdot  {\vec v}^\mu - i \frac{1}{2}v^\mu_S,
\, \, \, \, \, a^\mu = \frac{1}{2} {\vec \tau}\cdot  {\vec a}^\mu,
\\
f^+_{\mu \nu} &=& \partial_\mu \left(\frac{1}{2} {\vec \tau}\cdot  {\vec v}_{\nu} \right)
- \partial_\nu \left(\frac{1}{2}{\vec \tau}\cdot  {\vec v}_{\mu} \right), 
\, \, \, \, \,
v^{\mu \nu}_S = \frac{1}{2} (\partial^\mu v^\nu_S - \partial^\nu v^\mu_S),
\end{eqnarray}
where $g_A$ is the axial vector coupling constant $g_A = 1.267$, 
$m_N$ is the mean nucleon mass $m_N = (m_p + n_n)/2$, and
$\kappa_V$ and $\kappa_S$ are the isovector and isoscalar anomalous magnetic moments of the nucleon: 
$\kappa_V = 3.70589$ and $\kappa_S = -0.12019$.
$v^\mu$ is the velocity vector satisfying $v^2=1$. Assuming a non-relativistic limit, we have $v^\mu = (1, \vec{0})$,
which subsequently determines the spin operator $2S^\mu = (0, \vec{\sigma})$.
 $v^\mu_S$, $v^\mu_a$, and $a^\mu_a$ are the external isoscalar, 
isovector, and axial isovector fields, respectively,
and $\tau_a$ and $\sigma_i$ are the Pauli matrices for 
the isospin and spin, respectively.

The Lagrangian for the two-nucleon sector is given as
\begin{eqnarray}
{\cal L}_s &=& 
\sigma s^\dagger_a \left[ i v \cdot D + \frac{1}{4 m_N} 
\left[ (v\cdot D)^2 -D^2\right]
+ \Delta_s \right] s_a - y_s 
\left[ s^\dagger_a (N^T P^{(^1 {\rm S}_0)}_a N) + {\rm H.c.}\right],
\\
{\cal L}_t &=& \sigma t^\dagger_i \left[ i v \cdot D 
+ \frac{1}{4 m_N} \left[ (v\cdot D)^2 -D^2\right]
+ \Delta_t \right] t_i 
- y_t \left[ t^\dagger_i (N^T P^{(^3 {\rm S}_1)}_i N) + {\rm H.c.}\right],
\\
{\cal L}_{\rm int} &=& 
\frac{L_{1A}}{m_N \sqrt{\rho_d r_0}} \left[ a^i_a t^{i \dagger} s_a 
+ {\rm H.c.}\right]
+ \frac{L_1}{m_N \sqrt{\rho_d r_0}} \left[ B^i_a t^{i \dagger} s_a 
+ {\rm H.c.}\right]
\nonumber \\
& & + \frac{L_{2A}}{m_N \sqrt{\rho_d r_0}} 
\left[ 
i (\vec{D} a^0_a) \cdot \vec{t} s_a + {\rm H.c.} 
\right]\,,
\end{eqnarray}
where $D_\mu$ is the covariant derivative for the dibaryon fields,
$D_\mu = \partial_\mu -iC{\cal V}^{ext}_\mu$; 
${\cal V}^{ext}$ is the external vector field and $C$ is the charge 
operator of the dibaryon fields where $C=0$, 1, and 2 for the $nn$, $np$, and 
$pp$ channels, respectively.  
(See footnote 6 in Ref.~\cite{prc2005} as well.)
In addition, $\vec{B} = \nabla \times \vec{v}_a$, and 
$\sigma$ is the sign factor ($+1$ or $-1$), 
which is fixed so as to reproduce the amplitude 
in terms of the effective range expansion parameters.
$\Delta_s$ and $\Delta_t$ are defined as $\Delta_{s,t} \equiv m_{s, t} - 2m_N$,
 where $m_s$ and $m_t$ are the
masses of dibaryon fields in the $^1{\rm S}_0$ 
and $^3 {\rm S}_1$ states, respectively.
$\rho_d$ and $r_0$ are the effective ranges 
in the  $^1{\rm S}_0$ and $^3 {\rm S}_1$ states,
and the projection operators for each state are defined as
\begin{eqnarray}
P^{(^1 {\rm S}_0)}_a = \frac{1}{\sqrt{8}} \sigma_2 \tau_2 \tau_a\,,
\,\,\,\,\,
P^{(^3 {\rm S}_1)}_i = \frac{1}{\sqrt{8}} \sigma_2 \sigma_i \tau_2\,.
\end{eqnarray}
Moreover, $y_s$ and $y_t$ are determined from the effective range parameters, 
and we obtain 
\begin{eqnarray}
y^2_s = - \frac{8 \pi \sigma}{m^2_N r_0}, 
\,\,\,\,\,
y^2_t = - \frac{8 \pi \sigma}{m^2_N \rho_d}.
\end{eqnarray}
Determination of $L_{1A}$, $L_1$, and $L_{2A}$ will be discussed 
in Sec. III. 
 
\begin{figure}[tbp]
\begin{center}
\includegraphics[width=0.7\textwidth]{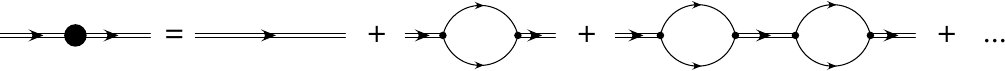}
\caption{
Feynman diagrams for dressed dibaryon propagator
for $np$ and $nn$ channels.
A single curve denotes the propagation of a nucleon 
while double lines with and without a filled circle denote
propagation of dressed and bare dibaryons, respectively. 
}
\label{fig1}
\end{center}
\end{figure}

\begin{figure}[tbp]
\begin{center}
\includegraphics[width=0.6\textwidth]{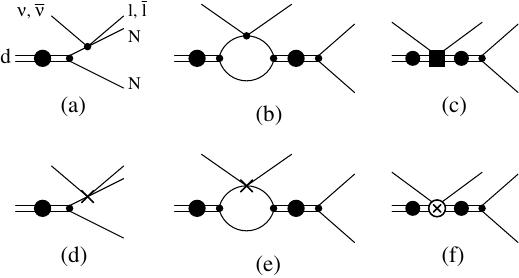}	
\caption{
Feynman diagrams for 
the $\nu d\to lNN$ and 
$\bar\nu d\to \bar{l} NN$ reactions 
without the Coulomb interaction between two nucleons
at LO (a)--(c) and NLO (d)--(f). 
Vertices with a dot (a cross) interacting with a lepton current 
are those 
at LO (NLO) 
for the single nucleon sector, 
and those with a filled box (a crossed circle)
are vertices at LO (NLO)  
for the double nucleon sector.
See the caption of Fig.~\ref{fig1} as well. 
}
\label{fig2}
\end{center}
\end{figure}

\begin{figure}[tbp]
\begin{center}
\includegraphics[width=0.7\textwidth]{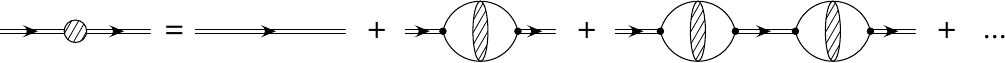}	
\caption{
Feynman diagrams for dressed dibaryon propagator
for the $pp$ channel.
Hatched ovals represent the non-perturbative Coulomb interaction between two protons,
which include no interaction diagram and all possible diagrams 
exchanging a potential photon up to infinite order. 
See the caption of Fig.~\ref{fig1} as well.
}
\label{fig3}
\end{center}
\end{figure}

\begin{figure}[tbp]
\begin{center}	
\includegraphics[width=0.6\textwidth]{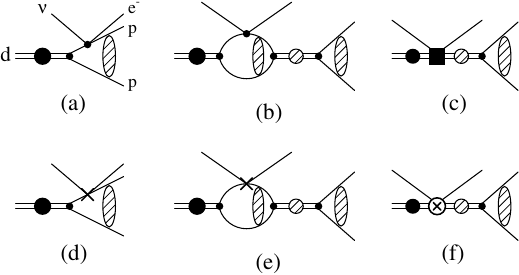}	
\caption{
Feynman diagrams for the $\nu d\to e^- pp$ 
reaction at LO (a)--(c) and NLO (d)--(f). 
See the captions in FIGs.~\ref{fig1}, \ref{fig2}, and \ref{fig3}
as well.
}
\label{fig4}
\end{center}
\end{figure}

Total cross section is calculated with the non-relativistic 
formula~\cite{hybrid}
\begin{eqnarray}
\sigma_{\nu d}(E_\nu) = \frac{1}{(2\pi)^3} \int dp \int dy
\frac{p^2 k'^2}{2E_\nu E'}
\frac{F(Z,E')}{k'/E' + (k'-E_\nu y)/(2m_N)} 
\frac{1}{2 S_d +1}
\sum_{\rm spin} |A_{(qq, np)}|^2,
\label{eq:cross_section}
\end{eqnarray}
with the condition for the energy-momentum conservation up to $1/m_N$ order,
\bea
m_d + E_\nu - E' -2m_N
-\frac{1}{m_N}\left[
p^2 
+ \frac14\left(
E_\nu^2 +k'^2 -E_\nu k' y
\right)
\right]=0\,,
\label{eq:energy_momentum_conservation}
\eea
where $m_d$ is the mass of the deuteron,
$E_\nu$ ($E'$) is the energy of the neutrino (lepton) 
in the initial (final) state,
$p$ is the magnitude of the relative three-momentum 
between the two nucleons in the final state, 
$k'$ is that of the outgoing lepton, and
$y$ is the cosine of angle between the incoming
and the outgoing leptons ($ y = \hat{k} \cdot \hat{k}'$). 
$F(Z, E')$ is the Fermi function taking into account 
the Coulomb interactions between the electron and the nucleons 
in the final state; 
its explicit form can be found in Ref.~\cite{morita1973}. 
Here, $S_d$ is the total spin of the deuteron, $S_d=1$.
We note that the expression of the total cross section 
in Eq.~(\ref{eq:cross_section}) is different from that in our
previous study~\cite{hybrid} by a factor of $1/(4E'E_\nu)$ 
because of different normalizations for the lepton fields. 
In addition, the fourth term, $2m_N$, in the left-hand side 
of Eq.~(\ref{eq:energy_momentum_conservation}) depends on 
the final two nucleon states:
$2m_N=2m_n,2m_p,m_n+m_p$ for the $nn$, $pp$, $np$ states,
where $m_n$ and $m_p$ are the masses of neutron and proton.
$A_{(nn, np)}$ are the transition amplitudes 
for the final $nn$ and $np$ channels, 
which are calculated from the 
Feynman diagrams shown in Figs.~\ref{fig1} and \ref{fig2},
and $A_{(pp)}$ is that for the final $pp$ channels
from Figs.~\ref{fig3} and \ref{fig4}. 
Diagrams (a)-(c) are the LO contributions, 
and (d)-(f) give the NLO contributions in Figs.~\ref{fig2}
and \ref{fig4}.

%
%
%
%
%
%

\subsection{Reaction amplitudes}

We write the amplitudes of charged current as
\begin{eqnarray}
A_{qq} &=& (-1)^{(1+\tau_q)/2} G'_F V_{ud} \left[
-{\cal A}_A^{1S0} + \tilde{\cal A}_V^{1S0} - \tilde{\cal A}_A^{1S0}
+\sum_{J=0}^2 \left(
{\cal A}_V^{3PJ} - {\cal A}_A^{3PJ}  
\right)
\right]\,,
\end{eqnarray}
where $q$ denotes neutron ($n$) or proton ($p$), and $\tau_{n (p)} =  -1 (1)$.
Leading order amplitudes are contributed from diagrams (a)-(c) 
in Figs. \ref{fig2} and \ref{fig4}, and 
they are represented in terms of the amplitudes ${\cal A}$ without tilde.
Letters $\tilde{\cal A}$ with tilde represent the amplitudes at NLO, 
and they are from diagrams (d)-(f) in Figs. \ref{fig2} and \ref{fig4}.
The superscripts and subscripts on the amplitudes denote
the partial waves and spin states ($^1S_0$, $^3S_1$, and $^3P_J$ states) 
for the final two nucleons and 
isovector vector ($V$) and axial-vector ($A$) nuclear currents, 
respectively. 

Neutral-current amplitudes are written as
\begin{eqnarray}
A_{np} &=& \frac{G'_F}{\sqrt2}\left\{
(1-2\sin^2\theta_W)\left[
\tilde{\cal A}_V^{1S0}
+ \sum_{J=0}^2 
{\cal A}_V^{3PJ} 
\right]
\right.
\nnb \\ && \left.
-\left(
{\cal A}_A^{1S0} + \tilde{\cal A}_A^{1S0} 
+ \sum_{J=0}^2 {\cal A}_A^{3PJ}
\right) -2\sin^2\theta_W\left[
{\cal A}_{VS}^{3S1} + \tilde{\cal A}_{VS}^{3S1}
\right]
\right\}\,, 
\end{eqnarray}
where the subscript ($VS$) denotes the isoscalar vector part of
the nuclear current.
Because the partial waves are orthogonal, we can write the squared amplitudes as
\begin{eqnarray}
|A_{qq}|^2 &=& G'^2_F V^2_{ud} \left[ \left|
-{\cal A}_A^{1S0} + \tilde{\cal A}_V^{1S0} - \tilde{\cal A}_A^{1S0} \right|^2
+ \sum^2_{J=0} \left|
{\cal A}_V^{3PJ} - {\cal A}_A^{3PJ} 
\right|^2
\right], \\
|A_{np}|^2 &=& \frac{G'^2_F}{2} 
\left\{ \left|{\cal A}_A^{1S0} + \tilde{\cal A}_A^{1S0} - (1 - 2 \sin^2 \theta_W) \tilde{\cal A}_V^{1S0}\right|^2
+ \left| 2 \sin^2 \theta_W ({\cal A}_{VS}^{3S1} + \tilde{\cal A}_{VS}^{3S1}) \right|^2 
\right.
\nonumber \\
&& +\left. 
\sum^2_{J=0} \left| {\cal A}_A^{3PJ} - (1- 2 \sin^2 \theta_W ) 
{\cal A}_V^{3PJ}
\right|^2 \right\}.
\end{eqnarray}

Amplitudes for S-wave states can be written as
\begin{eqnarray}
{\cal A}_A^{1S0} &=& 
\vec{\epsilon}_{(l)}\cdot \vec{\epsilon}_{(d)} X_A^{1S0}\,,
\\
\tilde{\cal A}_A^{1S0} &=& 
v\cdot\epsilon_{(l)}\hat{q}\cdot \vec{\epsilon}_{(d)} \tilde{X}_A^{1S0}\,,
\\
\tilde{\cal A}_V^{1S0} &=& 
i\hat{q}\cdot \left(\vec{\epsilon}_{(d)}\times\vec{\epsilon}_{(l)}\right) 
\tilde{X}_V^{1S0}\,,
\\
{\cal A}_{VS}^{3S1} &=& 
v\cdot\epsilon_{(l)}
\vec{\epsilon}^*\cdot\vec{\epsilon}_{(d)} X_{VS}^{3S1}\,,
\\
\tilde{\cal A}_{VS}^{3S1} &=& 
\hat{q}\cdot\vec{\epsilon}_{(l)}
\vec{\epsilon}^*\cdot \vec{\epsilon}_{(d)} \tilde{X}_{VS}^{3S1}\, ,
\end{eqnarray}
where $\vec{\epsilon}_{(d)}$ and $\vec{\epsilon}^*$ are spin polarization 
vectors of the incoming deuteron and the final two nucleon $^3S_1$ states,
respectively, and $\epsilon^{(l)}_\mu$ are the lepton currents.
In addition, $\hat{q}=\vec{q}/|\vec{q}|$, where $\vec{q}$ is the momentum
transfer between the lepton current and the nuclear current:
$\vec{q}=\vec{k}'-\vec{k}$.
$X$ and $\tilde{X}$ denote the LO and NLO contributions, respectively, 
and are given by
\begin{eqnarray}
X_A^{1S0} &=& 
2g_A\sqrt{
\frac{2\pi\gamma}{1-\gamma \rho_d}
}\left\{ \frac{}{}
\Gamma_{4(1)}^{(NN)}(p,q)
\right. \nnb \\ && \left.
+ \Gamma_{3(0)}^{(NN)}(p)D_s^{(NN)}(p)\left[
\Gamma_{3(1)}^{(NN)}(p,q) 
- \frac14(r^{(NN)}_0+\rho_d) 
+ \frac{l_{1A}}{2g_Am_N}
\right]
\right\}\,,
\\
\tilde{X}_A^{1S0} &=& 
\frac{2g_A}{m_N}\sqrt{
\frac{2\pi\gamma}{1-\gamma \rho_d}
}\left\{ \frac{}{} p\Gamma_{4(2)}^{(NN)}(p,q)
\right. \nnb \\ && \left.
+\Gamma_{3(0)}^{(NN)}(p) D_s^{(NN)}(p) q\left[
\Gamma_{3(2)}^{(NN)}(p,q) - \frac{1}{16}(r^{(NN)}_0+\rho_d)
+\frac{l_{2A}}{4g_A}
\right]
\right\}\,,
\\
\tilde{X}_V^{1S0} &=& 
-\frac{\mu_Vq}{m_N}\sqrt{
\frac{2\pi\gamma}{1-\gamma \rho_d}
}\left\{ \frac{}{}
\Gamma_{4(1)}^{(NN)}(p,q)
\right. \nnb \\ && \left.
+ \Gamma_{3(0)}^{(NN)}(p)D_s^{(NN)}(p)\left[
\Gamma_{3(1)}^{(NN)}(p,q) 
- \frac14(r^{(NN)}_0+\rho_d) 
+ \frac{l_{1}}{\mu_V}
\right]
\right\}\,,
\\
X_{VS}^{3S1} &=& 
-2\sqrt{
\frac{2\pi\gamma}{1-\gamma \rho_d}
}\left\{ \frac{}{}
\Gamma_{4(1)}^{(np)}(p,q)
+ \Gamma_{3(0)}^{(np)}(p)D_t^{(np)}(p)\left[
\Gamma_{3(1)}^{(np)}(p,q) - \frac12\rho_d
\right]
\right\}\,,
\label{eq;XVS3S1}
\\
\tilde{X}_{VS}^{3S1} 
&=& 
-\frac{2}{m_N}\sqrt{
\frac{2\pi\gamma}{1-\gamma\rho_d}
}\left\{
p\Gamma_{4(2)}^{(np)}(p,q)
+\Gamma_{3(0)}^{(np)}(p)D_t^{(np)}(p)q\left[
\Gamma_{3(2)}^{(np)}(p,q) -\frac{1}{8}\rho_d
\right]
\right\}\,,
\label{eq;tildeXVS3S1}
\end{eqnarray}
where $\gamma$ is the deuteron binding momentum,
$\gamma = \sqrt{2m_NB}$; $B$ is the deuteron binding
energy, and $\mu_V = 1+\kappa_V$.
The LECs $l_{1A}$, $l_{2A}$, and $l_1$ for the 
contact interactions are defined as 
\begin{eqnarray}
l_{1A} &=& 
L_{1A} 
+ \frac{1}{2}  m_N g_A (\rho_d + r^{(NN)}_0)\,, 
\\
l_{2A} &=&
L_{2A} 
+ \frac{1}{4} g_A (\rho_d + r^{(NN)}_0)\,,
\\
l_1 &=& L_1 
+ \frac{1}{4} \mu_V (\rho_d + r^{(NN)}_0)\,.
\end{eqnarray}
As discussed in Ref.~\cite{prc2005}, we separate the LECs $L_{1A}$, $L_{2A}$,
and $L_1$ into two parts: one consists of the effective range terms so as to
reproduce the result from the effective range theory, and the other consists of the 
unfixed constants $l_{1A}$, $l_{2A}$, and $l_1$, which may correspond to 
small corrections from a mechanism at high energy 
such as meson exchange currents.  
We note that LECs do not appear for the isoscalar vector parts 
in Eqs. (\ref{eq;XVS3S1}) and (\ref{eq;tildeXVS3S1}) due to conserved vector current (CVC). 
In addition, the sign for $l_{1A}$ above is opposite 
to that in Refs.~\cite{plb2008,nnfusion}.

$\Gamma_{4(1,2)}^{(NN)}(p,q)$ are four-point vertex functions for
the $NN=np,nn,pp$ channels,
$\Gamma_{3(0)}^{(NN)}(p)$ and $\Gamma_{3(1,2)}^{(NN)}(p,q)$ are three-point
vertex functions, 
and $D_s^{(NN)}(p)$ and $D_t^{(np)}(p)$ are dressed two-nucleon propagators
for spin singlet and spin triplet channels, respectively.
For the $np$ and $nn$ channels we have
\bea
\Gamma_{4(1)}^{(np,nn)}(p,q) &=&
\frac{1}{2pq}\ln\left(
\frac{\gamma^2+(p+\frac12q)^2}{\gamma^2+(p-\frac12q)^2 }
\right)\,,
\\
\Gamma_{4(2)}^{(np,nn)}(p,q) &=&
\frac{1}{pq}\left[
1
-\frac{\gamma^2+p^2+\frac14q^2}{2pq}\ln\left(
\frac{\gamma^2+(p+\frac12q)^2}{\gamma^2+(p-\frac12q)^2}
\right)
\right]\,,
\\
\Gamma_{3(1)}^{(np,nn)}(p,q) &=& \frac{1}{q}\left[
\frac{\pi}{2} - \arcsin\left(
\frac{\gamma^2+p^2-\frac14q^2}{\sqrt{
(\gamma^2+p^2+\frac14q^2)^2 - (pq)^2
}
}
\right)
+ \frac{i}{2}\ln\left(
\frac{\gamma^2+(p+\frac12q)^2}{\gamma^2+(p-\frac12q)^2}
\right)
\right]\,,
\nnb \\ &&
\\
\Gamma_{3(2)}^{(np,nn)}(p,q) &=&
-\frac{\gamma^2+p^2+\frac14q^2}{q^3}\left[
\frac{\pi}{2}-\arcsin\left(
\frac{\gamma^2+p^2-\frac14q^2}{\sqrt{
(\gamma^2+p^2+\frac14q^2)^2-(pq)^2
}
}
\right)
\right.
\nnb \\ && \left.
+ \frac{i}{2}\ln\left(
\frac{\gamma^2+(p+\frac12 q)^2}{\gamma^2+(p-\frac12q)^2}
\right)
\right]
+ \frac{1}{q^2}(\gamma+ip)\,,
\\
D_s^{(np,nn)}(p) &=& \frac{1}{
-\frac{1}{a_0^{(np,nn)}} + \frac12r_0^{(np,nn)}p^2 -ip
}\,,
\label{eq;Ds}
\\
D_t^{(np)}(p) &=& \frac{1}{
-\gamma + \frac12\rho_d(\gamma^2+p^2) -ip
}\,,
\label{eq;Dt}
\ \ \
\Gamma_{3(0)}^{(np,nn)}(p) = 1\,,
\eea
where $a_0^{(NN)}$ are scattering lengths of the $NN$ scattering 
in the $^1S_0$ channel.

For the $pp$ channel, we include the contribution from 
the nonperturbative Coulomb interaction in the amplitudes;
we follow the calculation method suggested 
by Ryberg {\it et al.}~\cite{rfhpprc2014},
in which Coulomb Green's functions are represented in the coordinate
space satisfying appropriate boundary conditions.\footnote{
Recently, this method was applied to a calculation of the $S_{E1}$ factor
of the $^{12}$C($\alpha$,$\gamma$)$^{16}$O reaction in an EFT~\cite{aprc2019}.
} 
Thus we have 
\bea
\Gamma_{4(1)}^{(pp)}(p,q) &=& \frac{e^{i\sigma_0}}{p}
\int_0^\infty drF_0(\eta,pr)j_0(\frac12qr)e^{-\gamma r}\,,
\\
\Gamma_{4(2)}^{(pp)}(p,q) &=& 
-e^{i\sigma_0} \frac{q}{p^2} \left[
\frac12\int_0^\infty dr F_0(\eta,pr)j_0(\frac12qr)e^{-\gamma r}
\right. \nnb \\ && \left.
-\frac{1}{q}\int_0^\infty dr F_0(\eta,pr)j_1(\frac12qr)\left(
\frac{1}{r} + \gamma
\right)e^{-\gamma r}
\right]
\,,
\\
\Gamma_{3(1)}^{(pp)}(p,q) &=&
C_\eta\int_0^\infty drH_0^+(\eta,pr)j_0(\frac12qr)e^{-\gamma r}\,,
\\
\Gamma_{3(2)}^{(pp)}(p,q) &=&
- C_\eta \left[
\frac12\int_0^\infty dr H_0^+(\eta,pr)j_0(\frac12qr)e^{-\gamma r}
\right. \nnb \\ && \left.
-\frac{1}{q}\int_0^\infty dr H_0^+(\eta, pr) j_1(\frac12qr)\left(
\frac{1}{r}+\gamma
\right)e^{-\gamma r}
\right]\,,
\\
D_s^{(pp)}(p) &=& \frac{1}{
-\frac{1}{a_0^{(pp)}}+\frac12r_0^{(pp)}p^2 + 2\kappa H(\eta)
}\,,
\label{eq;Ds_pp}
\ \ \
\Gamma_{3(0)}^{(pp)}(p) = e^{i\sigma_0}C_\eta\,,
\\
H(\eta) &=& \psi(i\eta) + \frac{1}{2i\eta} -\ln(i\eta)\,,
\ \ \
C_\eta = \sqrt{
\frac{2\pi\eta}{e^{2\pi\eta}-1}
}\,,
\ \ \ 
e^{i\sigma_0} = \sqrt{
\frac{\Gamma(1+i\eta)}{\Gamma(1-i\eta)}
}\,,
\eea
where $H_l^+(\eta,\rho)= G_l(\eta,\rho)+iF_l(\eta,\rho)$;
$F_l(\eta,\rho)$ and $G_l(\eta,\rho)$
are regular and irregular Coulomb wave functions, respectively,
and $j_l(x)$ are the spherical Bessel functions.
$\psi(z)$ is the digamma function with $\eta = \kappa/p$;
$\kappa = \alpha_E m_p/2$ where $\alpha_E$ is the fine-structure constant.
The $r$-space integrations above can be carried out analytically.
The derivation and expression of those integrals are presented in 
Appendix A.

In a similar manner we write the amplitudes in the $P$-wave states as
\begin{eqnarray}
{\cal A}_V^{3P0} &=& 
v\cdot \epsilon_{(l)} \hat{q}\cdot \vec{\epsilon}_{(d)}
X_V^{3P0}\,,
\\
{\cal A}_A^{3P0} &=& 
i\hat{q}\cdot\left(\vec{\epsilon}_{(l)}\times \vec{\epsilon}_{(d)}
\right) X_A^{3P0}\,,
\\
{\cal A}_V^{3P1} &=& iv\cdot\epsilon_{(l)} 
\vec{\epsilon}^*\cdot\left(
\hat{q}\times \vec{\epsilon}_{(d)}
\right) X_V^{3P1}\,,
\\
{\cal A}_A^{3P1} &=& \left(
\vec{\epsilon}^*\cdot \vec{\epsilon}_{(l)}\hat{q}\cdot\vec{\epsilon}_{(d)}
-\vec{\epsilon}^*\cdot\vec{\epsilon}_{(d)}\hat{q}\cdot\vec{\epsilon}_{(l)}
\right) X_A^{3P1}\,,
\\
{\cal A}_V^{3P2} &=& v\cdot \epsilon_{(l)}
\epsilon^{ij*} \hat{q}^i \epsilon_{(d)}^j X_V^{3P2}\,,
\\
{\cal A}_A^{3P2} &=& i\epsilon^{ij*} \hat{q}^i \epsilon^{jkl}
\epsilon_{(l)}^k\epsilon_{(d)}^l X_A^{3P2}\,,
\end{eqnarray}
with
\begin{eqnarray}
X_V^{3P0} &=& 
- 2\sqrt{
\frac{2\pi\gamma}{1-\gamma \rho_d}
} \Gamma_{4(3)}^{(NN)}(p,q)\,,
\\ 
X^{3P1}_V &=& \sqrt{\frac32}X_V^{3P0}\,,
\ \ \
X^{3P2}_V = \sqrt{3} X_V^{3P0}\,,
\\
X^{3PJ}_A &=& -g_AX^{3PJ}_V \,, 
\end{eqnarray}
where $J=0,1,2$, and $\epsilon_i^*$ and $\epsilon_{ij}^*$ are a vector and 
a symmetric tensor representing the final two nucleons for $J=1$ and 2 states,
respectively.
For the $np$ and $nn$ channels, 
$\Gamma_{4(3)}^{(np,nn)}(p,q) = \Gamma_{4(2)}^{(np,nn)}(p,q)$, 
and for the $pp$ channel we have 
\bea
\Gamma_{4(3)}^{(pp)}(p,q) &=& - \frac{e^{i\sigma_1}}{p}
\int_0^\infty dr F_1(\eta,pr)j_1(\frac12qr)e^{-\gamma r}\,,
\eea 
with
\bea
e^{i\sigma_1} = \sqrt{
\frac{\Gamma(2+i\eta)}{\Gamma(2-i\eta)}
}\,.
\eea
An analytic expression of the vertex function $\Gamma_{4(3)}^{(pp)}(p,q)$ 
is presented in Appendix A as well. 

Summing over spins in the initial and final states,
we obtain the result
\begin{eqnarray}
&&
\sum_{\rm spins}|A_{qq}|^2 = 8G_F'^2 V^2_{ud} \left\{
(3E'E- \vec{k}'\cdot\vec{k})\left|
X_A^{1S0}
\right|^2
\right. 
-\left(
E'\hat{q}\cdot\vec{k} + E\hat{q}\cdot\vec{k}'
\right)\left[
X_A^{1S0 *} \tilde{X}_A^{1S0}
+ \tilde{X}_A^{1S0 *} X_A^{1S0}
\right]
\nnb
\\ 
&&
\mp 2\left(
E'\hat{q}\cdot\vec{k} - E\hat{q}\cdot\vec{k}'
\right) \left[
X_A^{1S0*} \tilde{X}_V^{1S0}
+ \tilde{X}_V^{1S0*} X_A^{1S0}
\right]
\nnb \\ && 
+ (E'E+\vec{k}'\cdot\vec{k}) \left|
\tilde{X}_A^{1S0} \right|^2
+2(E'E-\hat{q}\cdot\vec{k}'\hat{q}\cdot\vec{k})
\left|
\tilde{X}_V^{1S0} \right|^2
\nnb \\ 
&& +(E'E+\vec{k}'\cdot \vec{k})\left[
\left|
X_V^{3P0}
\right|^2
+ 2\left|
X_V^{3P1}
\right|^2
+ \frac53\left|
X_V^{3P2}
\right|^2
\right]
\nnb \\ 
&& + 2(E'E-\hat{q}\cdot\vec{k}'\hat{q}\cdot\vec{k})
\left|
X_A^{3P0}
\right|^2
+ 2(2E'E - 2\vec{k}'\cdot\vec{k}-\hat{q}\cdot\vec{k}'\hat{q}\cdot\vec{k})
\left|
X_A^{3P1}
\right|^2
\nnb \\ && \left.
+ 4 (20E'E - 6 \vec{k}'\cdot\vec{k}+2\hat{q}\cdot\vec{k}'\hat{q}\cdot\vec{k})
\left|
X_A^{3P2}
\right|^2
\right\}\,,
\\
&&
\sum_{\rm spins}|A_{np}|^2 = 4G_F'^2\left\{
(3E'E- \vec{k}'\cdot\vec{k})\left|
X_A^{1S0}
\right|^2
-\left(
E'\hat{q}\cdot\vec{k} + E\hat{q}\cdot\vec{k}'
\right)\left[
X_A^{1S0*} \tilde{X}_A^{1S0}
+ \tilde{X}_A^{1S0*} X_A^{1S0}
\right]
\right.
\nnb
\\ &&
\mp 2\left(
E'\hat{q}\cdot\vec{k} - E\hat{q}\cdot\vec{k}'
\right) \left(
1-2\sin^2\theta_W
\right)\left[
X_A^{1S0*} \tilde{X}_V^{1S0}
+ \tilde{X}_V^{1S0*} X_A^{1S0}
\right]
+ (E'E+\vec{k}'\cdot\vec{k}) \left|
\tilde{X}_A^{1S0}
\right|^2
\nnb \\ && 
+2(E'E-\hat{q}\cdot\vec{k}'\hat{q}\cdot\vec{k})
\left(
1-2\sin^2\theta_W
\right)^2
\left|
\tilde{X}_V^{1S0}
\right|^2
+ 12 \sin^4 \theta_W (E'E+\vec{k}'\cdot\vec{k})\left|
X_{VS}^{3S1}
\right|^2
\nnb \\ && 
-12 \sin^4 \theta_W \left(E'\hat{q}\cdot\vec{k}+E\hat{q}\cdot\vec{k}'
\right)
\left[
X_{VS}^{3S1*} \tilde{X}_{VS}^{3S1} +  \tilde{X}_{VS}^{3S1*} X_{VS}^{3S1}
\right]
\nnb \\ && 
+ 12 \sin^4 \theta_W \left(
E'E - \vec{k}'\cdot\vec{k} -2 \hat{q}\cdot\vec{k}'\hat{q}\cdot\vec{k}
\right)\left|
\tilde{X}_{VS}^{3S1}
\right|^2
\nnb \\ && 
+(E'E+\vec{k}'\cdot \vec{k})\left(
1-2\sin^2\theta_W
\right)^2
\left[
\left|
X_V^{3P0}
\right|^2
+ 2\left|
X_V^{3P1}
\right|^2
+ \frac53\left|
X_V^{3P2}
\right|^2
\right]
\nnb \\ && + 2(E'E-\hat{q}\cdot\vec{k}'\hat{q}\cdot\vec{k})
\left|
X_A^{3P0}
\right|^2
+ 2(2E'E - 2\vec{k}'\cdot\vec{k}-\hat{q}\cdot\vec{k}'\hat{q}\cdot\vec{k})
\left|
X_A^{3P1}
\right|^2
\nnb \\ && \left.
+ 4 (20E'E - 6 \vec{k}'\cdot\vec{k}+2\hat{q}\cdot\vec{k}'\hat{q}\cdot\vec{k})
\left|
X_A^{3P2}
\right|^2
\right\}\,,
\end{eqnarray}
where we have used the spin summation relations presented in 
Appendix B.

\section{Numerical results}


First we specify the values of LECs.
Low-energy constant $l_1$ is fitted to the total cross section of the radiative neutron-proton 
capture process at threshold \cite{prc2005}.
Data that can constrain the numerical value of $l_{1A}$ 
directly from experiment are not available;
we employ two values of $l_{1A}$ reported 
in the previous works~\cite{plb2008,nnfusion}.
In the work on proton-proton fusion \cite{plb2008}, 
$l_{1A}$ is determined 
by using a ratio of the two-body amplitude to the one-body one 
for $pp$ fusion (reported in Eq.~(28) of Ref.~\cite{petal-prc03})
calculated in the hybrid EFT approach, where 
the strength of the two-body operator is constrained 
by the tritium lifetime; it gives $l_{1A} = 0.50 \pm 0.03$.
Another way to fix $l_{1A}$ is proposed in the work on neutron-neutron 
fusion \cite{nnfusion},
in which a value of $l_{1A}$ is constrained by the cross section 
of the $\nu d$ reaction for the $\nu d\to e^+nn$ channel
reported in Refs.~\cite{netal-npa,hybrid};
it gives $l_{1A} = 0.33 \pm 0.03$.
Since a robust way to fix the value of $l_{1A}$ is absent at present,
$l_{1A}$ is a source of uncertainty in the theoretical prediction.
Without any priority to a specific value, 
we use both values $l_{1A}=0.33$ and $0.50$ 
in the calculation of the total cross sections.~\footnote{
Recently, De-Leon, Platter, and Gazit reported a study of 
the tritium $\beta$ decay in pionless EFT and fitted 
a value of $l_{1A}$ to the tritium lifetime~\cite{dlpg-prc}. 
One should note, however, that the expansion scheme for the 
effective range terms are different between that work and the 
present one. We discuss that issue in Appendix C. 
} 
The Lagrangian term proportional to LEC $l_{2A}$ has 
a derivative one order higher than 
the terms proportional to $l_1$ and $l_{1A}$. 
Since the perturbative expansion is performed with respect to either energy or momentum
of the particles, higher order derivatives will be suppressed compared to the lower order ones.
In addition, since experimental data that can constrain $l_{2A}$ are not available,
we assume $l_{2A}=0$ for simplicity.~\footnote{
A model calculation including meson exchange currents 
suggests that the $A_0$ contribution is tiny. 
See Table III in Ref.~\cite{snpa}.
}
Scattering lengths $a^{(NN)}_0$ are taking the values $-23.73$~fm, $-18.50$~fm and $-7.82$~fm for 
$np$, $nn$ and $pp$ states, respectively, and effective ranges $r^{(NN)}_0$ are 
2.73~fm, 2.83~fm and 2.78~fm for the $np$, $nn$, and $pp$ states, respectively.

\begin{table}
\begin{center}
\begin{tabular}{ccccc}\hline\hline 
$E_\nu$ & $\nu np$ & $\bar{\nu} np$ & $e^+ n n$ & $e^- pp$ \\ \hline
2 & 0 & 0 & 0 & 0.003665 \\
3 & 0.003366 & 0.003330 & 0 & 0.04706 \\
4 & 0.03070 & 0.03021 & 0 & 0.1567 \\
5 & 0.09495 & 0.09291 & 0.02841 & 0.3459 \\
6 & 0.2017 & 0.1962 & 0.1192 & 0.6230 \\
7 & 0.3540 & 0.3425 & 0.2823 & 0.9936 \\
8 & 0.5540 & 0.5329 & 0.5225 & 1.462 \\
9 & 0.8029 & 0.7679 & 0.8422 & 2.032 \\
10 & 1.102 & 1.048 & 1.242 & 2.705 \\
11 & 1.452 & 1.373 & 1.722 & 3.485 \\
12 & 1.853 & 1.742 & 2.282 & 4.374 \\
13 & 2.307 & 2.157 & 2.922 & 5.374 \\
14 & 2.813 & 2.616 & 3.640 & 6.488 \\
15 & 3.373 & 3.119 & 4.436 & 7.716 \\
16 & 3.987 & 3.667 & 5.309 & 9.062 \\
17 & 4.656 & 4.259 & 6.259 & 10.53 \\
18 & 5.380 & 4.895 & 7.285 & 12.12 \\
19 & 6.161 & 5.576 & 8.386 & 13.83 \\
20 & 6.998 & 6.301 & 9.563 & 15.67 \\
\hline
\end{tabular}
\end{center}
\caption{Total cross sections of neutrino-deuteron scattering as functions of the incident neutrino energy.
Reaction types are denoted by the particles in the final states.
Incident neutrino energy $E_\nu$ is in MeV, and the total cross sections are in units of $10^{-42} {\rm cm}^2$.}
\label{tab:total}
\end{table}

\begin{figure}[tbp]
\begin{center}
\includegraphics[width=0.7\textwidth]{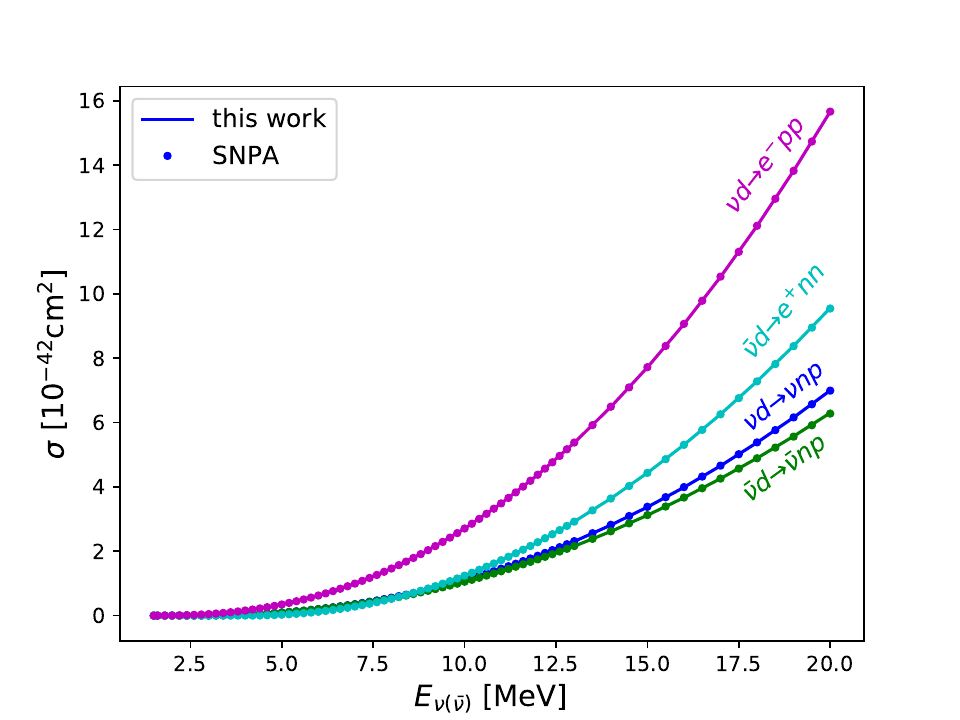} \phantom{xxx}
\caption{
Color online. Total cross sections (cm$^2$) as functions of the incident neutrino energy $E_\nu$.
Solid lines denote the result of this work for $\nu np$ (blue), 
$\bar{\nu} n p$ (green), $e^+ n n$ (cyan) and $e^- p p$ (magenta). 
Dots present the result of SNPA \cite{snpa}. 
}
\label{fig5}
\end{center}
\end{figure}

Table~\ref{tab:total} shows the total cross sections in units of $10^{-42}$~cm$^2$.
For the $np$ and $pp$ final states, results with $l_{1A}=0.50$ are presented, and the results for the $nn$ state are
with $l_{1A}=0.33$. The general trend is a monotonic increase with energy,
and the rate is larger in charged-current processes than in neutral ones.
For a better overall view, results in Table~\ref{tab:total} are plotted in Fig.~\ref{fig5}.
As a benchmark for comparison with other theories, we include the result of Ref.~\cite{snpa},
which is labeled SNPA meaning standard nuclear physics approach.
In the SNPA, initial- and final-state wave functions are the solutions of Schr\"{o}dinger equations with 
modern phenomenological $NN$ potentials, and transition operators consist of one-body impulse approximation
and two-body meson-exchange currents.
In our work we use partial wave expansion of the final state up to $P$-wave, but  
the SNPA results include states up to $J=6$.
Despite the huge differences in the basic formalism of the two theories on one hand, and partial waves
in the final state on the other hand, predictions of the two works agree remarkably well.
Recent EFT works \cite{butler,baroni} show agreement with SNPA within the order of 1\%.
A refined comparison is in order next.

\begin{table}
\begin{center}
\begin{tabular}{ccccccccc}\hline\hline 
$E_\nu$ & \multicolumn{2}{c}{$\nu np$} & \multicolumn{2}{c}{$\bar{\nu}np$} 
& \multicolumn{2}{c}{$e^+ n n$} 
& \multicolumn{2}{c}{$e^- pp$} \\ \hline
$l_{1A}$ &\phantom{x} 0.33 &\phantom{xx} 0.50 \,\,\,\,\ & 
\phantom{x} 0.33 & \phantom{xx} 0.50 \,\,\,\,\, & 
\phantom{x}0.33 & \phantom{xx} 0.50 \,\,\,\,\, & 
\phantom{x} 0.33 & \phantom{xx} 0.50
 \,\,\,\,\, \\ \hline
2 & - & -  & - & -  & 
- & -  & $-0.77$ & $-0.13$  \\
3 & $-1.79$ & $-1.06$ & $-1.81$ & $-1.07$ & 
- & - & $-0.82$ & $-0.12$\\
4 & $-1.69$ & $-0.94$ & $-1.69$ & $-0.92$  & 
- & - & $-0.88$ & $-0.13$ \\
5 & $-1.60$ & $-0.78$ & $-1.61$ & $-0.78$ & 
0.46 & 0.92 & $-0.89$ & $-0.11$  \\
6 & $-1.53$ & $-0.69$ & $-1.54$ & $-0.71$ & 
0.18 & 1.01 & $-0.90$ & $-0.10$  \\
7 & $-1.47$ & $-0.62$ & $-1.78$ & $-0.61$ & 
0.10 & 0.92 & $-0.92$ & $-0.10$  \\
8 & $-1.44$ & $-0.56$ & $-1.44$ & $-0.54$ & 
0.10 & 0.96 & $-0.97$ & $-0.13$ \\
9 & $-1.40$ & $-0.51$ & $-1.40$ & $-0.51$ & 
0.07 & 0.94 & $-0.96$ & $-0.12$  \\
10& $-1.37$ & $-0.47$ & $-1.42$ & $-0.48$ & 
0.01 & 0.90 & $-0.96$ & $-0.10$  \\
11& $-1.36$ & $-0.44$ & $-1.33$ & $-0.38$ &
0.02 & 0.93 & $-0.98$ & $-0.11$  \\
12& $-1.31$ & $-0.38$ & $-1.28$ & $-0.31$ & 
0.02 & 0.95 & $-0.99$ & $-0.11$ \\
13& $-1.26$ & $-0.32$ & $-1.26$ & $-0.28$ & 
$-0.00$ & 0.94 & $-1.00$ & $-0.11$ \\
14& $-1.26$ & $-0.31$ & $-1.22$ & $-0.23$ &
$-0.02$ & 0.94 & $-1.01$ & $-0.11$  \\
15& $-1.22$ & $-0.26$ & $-1.18$ & $-0.18$ &
$-0.02$ & 0.95 & $-1.01$ & $-0.10$ \\
16& $-1.16$ & $-0.20$ & $-1.12$ & $-0.11$ &
$-0.01$ & 0.97 & $-1.01$ & $-0.10$  \\
17& $-1.12$ & $-0.15$ & $-1.04$ & $-0.02$ &
0.02 & 1.01 & $-1.03$ & $-0.11$ \\
18& $-1.06$ & $-0.09$ & $-0.94$ & $0.09$ & 
0.04 & 1.04 & $-0.95$ & $-0.03$  \\
19& $-1.00$ & $-0.02$ & $-0.84$ & $0.20$ & 
0.09 &  1.11 & $-0.93$ & $-0.01$  \\
20& $-0.93$ & $0.06$ & $-0.72$ & $0.32$ &
0.16 &  1.19 & $-0.94$ & $-0.01$ \\
\hline
\end{tabular}
\end{center}
\caption{Differences of between our result and those using SNPA. Difference is calculated in \% as
(this work $-$ other)/other$\times 100$.}
\label{tab:difference}
\end{table}

In Table~\ref{tab:difference}, we show the difference between our work and SNPA.
For each reaction channel, we calculate the differences with both $l_{1A}=0.33$ and $0.50$. 
From the resulting differences, one can deduce that the total cross is larger with $l_{1A}=0.50$ than with 0.33 
for all reactions and energies.
$l_{1A}=0.50$ is adopted from the work on $pp$ fusion, and it gives better agreement with SNPA than $l_{1A}=0.33$
for the $e^-pp$ reaction channel.
On the other hand, the total cross section of the $e^+ nn$ channel is closer to SNPA with $l_{1A}=0.33$ than 0.50.
It turns out that $l_{1A}=0.50$ gives better agreement with SNPA than 0.33 for the neutral-current processes.
The gap between the differences of $l_{1A}=0.33$ and 0.50 is in the interval 0.7$-$1.1\%,
and it is weakly dependent on the reaction type and energy.
We take this as a theoretical uncertainty in this work originating from the LEC $l_{1A}$.
There are other sources of uncertainties such as
truncation at NLO in the perturbative expansion and $P$-wave approximation in the partial wave expansion.
Discussions of these uncertainties will be presented in the following paragraphs.

\begin{table}
\begin{center}
\begin{tabular}{ccccccccc} \hline\hline
 & \multicolumn{2}{c}{LO/total} & & \multicolumn{2}{c}{S-wave/total} & & \multicolumn{2}{c}{S-wave/EFT*} \\ \hline
$E_\nu$ & $\nu n p$ & $e^- pp$ & & $\nu n p$ & $e^- pp$ & & $\nu n p$ & $e^- pp$ \\ \hline
5  & 0.9886 & 0.9777 & &  0.9999 & 0.9997 & & - & - \\
10 & 0.9745 & 0.9525 & & 0.9975 & 0.9963 & & 0.994 & 0.999 \\
15 & 0.9609 & 0.9288 & & 0.9902 & 0.9875 & & 0.997 & 1.003 \\
20 & 0.9482 & 0.9070 & & 0.9764 & 0.9723 & & 1.000 & 1.007 \\
\hline
\end{tabular}
\end{center}
\caption{Ratios of the LO contribution to the total (LO/total), S-wave contribution to the total (S-wave/total),
and S-wave contribution in our work to that in EFT* \cite{hybrid} (S-wave/EFT*).}
\label{LOS}
\end{table}

The relative contribution of LO amplitudes to the total cross section is 
shown in the column  `LO/total' for the $\nu np$ and $e^- pp$ reactions
in Table \ref{LOS}.
At energies close to threshold, LO takes 98$-$99\% of the total.
Its portion decreases monotonically as energy increases,
and it reaches $91-95$\% at 20 MeV.
This result satisfies the general behavior of perturbation theory:
(i) lower orders dominate at low energies,
(ii) contributions of higher orders increase as the energy increases,
and (iii) higher order contributions should be sufficiently smaller than those of lower orders 
in the considered energy range.

In order to check the validity of partial wave approximation in the final state, 
we consider the contribution of $S$-wave states.
The column  `$S$-wave/total' in Table~\ref{LOS} shows the ratio of 
cross section from the $S$-wave contribution to that of the full contribution.
In the solar neutrino energy region, the total result is absolutely dominated by the $S$ waves.
However, though it is small, the contribution of higher partial waves increases with energy.
In our result the contribution of P waves at $E_\nu=5$~MeV is 0.03\% at most, and it increases to 2$-$3\% at $E_\nu=20$~MeV.
In the work using SNPA \cite{snpa}, the authors performed a similar analysis of the $S$-wave contribution.
At $E_\nu=20$~MeV, the proportions of the $^1{\rm S}_0$ state to the net value are obtained as
0.972 and 0.964 for $\nu n p$ and $e^- pp$ reactions, respectively.
These values are smaller than our results only by 0.004 and 0.008, so the two very different models
give consistent predictions about the role of S-wave states in the final state.
The $P$-wave contribution was also calculated separately in \cite{snpa}.
At $E_\nu=20$~MeV, the sum of $S$- and $P$-wave contribution is 99.9\% of the total result 
which includes the partial waves to $J=6$.
Therefore contributions from partial waves higher than the $D$ state will not be a source of uncertainty at the order of 1\%
for the solar neutrino energies.  

In Ref.~\cite{hybrid}, the authors investigated the same problem with a pionful EFT.
In that work, final state wave functions contain only the $S$-wave states.
In the column denoted `S-wave/EFT*' in Table~\ref{LOS}, we present the ratio of $S$-wave contributions in our work to those of EFT*.
The two theories agree in the range 0.994$-$1.007, so in practice the two theories predict equivalent results for the total cross section.

\section{Summary}

We considered the breakup of deuterons by neutrinos 
and antineutrinos at solar neutrino energies.
We calculated the total cross section of the neutral-current reactions
$\nu d\to \nu n p$, $\bar{\nu} d\to \bar{\nu} n p$, 
and the charged-current ones
$\bar{\nu} d\to e^+ n n$, $\nu d\to e^- p p$ 
in the framework of a pionless effective field
theory with dibaryon fields up to the next-to-leading order.
We included the Coulomb interaction between two protons nonperturbatively
while analytic expressions of the amplitudes were obtained.
We estimated the uncertainty of our theory by comparing 
our result to those obtained from
various theories and models.

In the comparison to the work that employs phenomenological nucleon-nucleon potential models,
the contribution of the $S$-wave state in the final state agrees well with the result of \cite{snpa},
and it is confirmed that the truncation at $P$ waves in the final state wave functions is a good approximation.
The main source of the uncertainty was identified as a low-energy constant $l_{1A}$ which determines the
strength of the axial four-nucleon contact interactions.
The low-energy constant determined from available experimental data gives uncertainties about 1\% or less.

We compared our results of the $S$-wave contribution with those obtained from a pionful effective field theory 
in which expansion is performed up to next-to-next-to-next-to-leading order \cite{hybrid}.
We found that the two theories predict practically identical results, with a difference of about 0.7\% at most.

Convergence of the theory was checked by isolating the leading order contribution from the total.
At low energies, leading order contributes almost 100\% of the total, and its portion decreases as the energy increases.
For the $\nu np$ reaction, leading order takes 95\% of the total at $E_\nu = 20$~MeV.
This ratio of the S-wave contribution is similar to the ratio of next-to-next-to-leading order to leading order 
obtained in a pionless effective field theory \cite{butler}.
This comparison demonstrates that the rate of convergence could be improved by the introduction of dibaryon
fields.

Our result underestimates the result of a benchmark calculation \cite{snpa} by about 1\%.
On the other hand, another pionless effective field theory \cite{butler} gives a result very similar to \cite{snpa},
and a chiral perturbation theory \cite{baroni} 
obtains results larger than \cite{snpa} by about 1\%.
Therefore one can conclude that the uncertainty one can obtain from the state-of-the-art theories is about
2$-$3\% in the solar neutrino energy range. 
Precise determination of LECs from experiment will be important to reduce the theoretical uncertainty.

\section*{Acknowledgments}

S.I.A. would like to thank K. Kubodera for useful discussions
and collaboration at the early stage of the present work, started 
in 2005.
The work of S.I.A. was supported by 
the Basic Science Research Programs through the National Research
Foundation of Korea 
(2016R1D1A1B03930122 and 2019R1F1A1040362).
The work of Y.H.S. was supported by the Rare Isotope Science Project 
of Institute for Basic Science, funded by Ministry of Science, 
ICT and Future Planning and by National Research Foundation of Korea 
(2013M7A1A1075764).
The work of C.H.H. was supported by the National Research Foundation of Korea (NRF)
with a grant funded by the Korea government (MSIT) (No. 2018R1A5A1025563).

\section*{Appendix A}

Analytic expression of the four-point vertex functions
$\Gamma_{4(1)}^{(pp)}(p,q)$ and $\Gamma_{4(3)}^{(pp)}(p,q)$ 
are calculated by using formula~\cite{NISTHB},
\bea
F_l(\eta,\rho) &=& C_l(\eta) \rho^{l+1} e^{\mp i\rho}
M(l+1\mp i\eta,2l+2,\pm 2i\rho)\,,
\label{eq;Fl}
\eea
where $M(a,b,z)$ is the Kummer function and
\bea
C_l(\eta) &=& \frac{2^le^{-\frac{\pi}{2}\eta}
|\Gamma(l+1+i\eta)|}{
(2l+1)!
}
=
\frac{2^l}{(2+1)!}\sqrt{
\frac{2\pi\eta}{e^{2\pi\eta}-1}\prod_{k=1}^l(\eta^2+k^2)
}\,.
\eea
The expressions for the ($\pm$ or $\mp$)
signs in Eq.~(\ref{eq;Fl}) are identical
because of the relation $M(a,b,z) = e^zM(b-a,b,-z)$.
Using another relation,
\bea
\int_0^\infty e^{-zt}t^{b-1}M(a,c,kt)dt
&=& \Gamma(b)z^{-b}{}_2F_1(a,b,c;k/z) \,,
\eea
with Re$[b]>0$, Re$[z]>\textrm{max}(\rm{Re}[k],0)$,
where ${}_2F_1(a,b,c;z)$ is a hypergeometric function,
we have
\bea
\Gamma_{4(1)}^{(pp)}(p,q) &=&  \frac{e^{i\sigma_0}}{p}
\int_0^\infty dr F_0(\eta,pr)j_0(\frac12qr)e^{-\gamma r} 
\nnb \\
&=& i\frac{e^{i\sigma_0}}{q}C_\eta\left[
\frac{1}{\gamma + i (p+\frac12q)}{}_2F_1\left(
1-i\eta, 1, 2; \frac{2ip}{\gamma + i(p+\frac12q)}
\right)
\right.
\nnb \\ && \left.
- \frac{1}{\gamma + i (p-\frac12q)}{}_2F_1\left(
1-i\eta, 1, 2; \frac{2ip}{\gamma + i(p-\frac12q)}
\right)
\right]\,,
\\
\Gamma_{4(3)}^{(pp)}(p,q) &=& - \frac{e^{i\sigma_1}}{p}
\int_0^\infty dr F_1(\eta,pr)j_1(\frac12qr)e^{-\gamma r} 
\nnb \\ &=&
-\frac13 p e^{i\sigma_1}\sqrt{
\frac{2\pi\eta(1+\eta^2)}{e^{2\pi\eta}-1}
}
\nnb \\ && \times \left\{
\frac{2i}{q^2}\left[
\frac{1}{\gamma + i(p+\frac12q)}{}_2F_1\left(
2-i\eta,1,4;\frac{2ip}{\gamma + i(p+\frac12q)}
\right)
\right.\right.
\nnb \\ && \left.
-\frac{1}{\gamma+i(p-\frac12q)}{}_2F_1\left(
2-i\eta,1,4;\frac{2ip}{\gamma + i(p-\frac12q)}
\right)
\right]
\nnb \\ &&
-\frac{1}{q}\left[
\frac{1}{[\gamma + i(p+\frac12q)]^2}
{}_2F_1\left(
2-i\eta,2,4;\frac{2ip}{\gamma+i(p+\frac12q)}
\right)
\right.
\nnb \\ && \left. \left.
+ \frac{1}{[\gamma+i(p-\frac12q)]^2}
{}_2F_1\left(
2-i\eta,2,4;\frac{2ip}{\gamma + i(p-\frac12q)}
\right)
\right]
\right\}\,.
\eea

The four point vertex $\Gamma_{4(2)}^{(pp)}(p,q)$ can be represented 
by using the result for $\Gamma_{4(1)}^{(pp)}(p,q)$ and an integration,
\bea
I_3 &=& \int_0^\infty dr F_0(\eta,pr)j_1(\frac12qr)\left(
\frac{1}{r} + \gamma
\right)e^{-\gamma r}\,,
\eea
and we have
\bea
I_3 &=& pC_\eta\sum_{n=1}^\infty (-1)^n\left[
\frac{1}{(2n+1)!} - \frac{1}{(2n)!}
\right]\left(
\frac{q}{2}
\right)^{2n-1}
\nnb \\ && \times \left[
\frac{\gamma\Gamma(2n+1)}{(\gamma+ip)^{2n+1}}
{}_2F_1\left(
1-i\eta,2n+1,2; \frac{2ip}{\gamma+ip}
\right)
\right.
\nnb \\ && \left.
+\frac{\Gamma(2n)}{(\gamma+ip)^{2n}}
{}_2F_1\left(
1-i\eta,2n,2; \frac{2ip}{\gamma+ip}
\right)
\right]\,.
\eea

An analytic expression for the three-point vertex function
$\Gamma_{3(1)}^{(pp)}(p,q)$ is obtained by using formulas,
\bea
H_l^\pm(\eta,\rho) &=&  (\mp)^le^{\frac{\pi}{2}\eta\pm i\sigma_l}
W_{\mp i\eta,l+\frac12}(\mp 2i\rho)\,,
\\
\int_0^\infty dt
e^{-pt}
t^{\nu-1}W_{\kappa,\mu}(at)
&=&
\frac{\Gamma(\mu+\nu+\frac12)\Gamma(\nu-\mu+\frac12)a^{\nu+\frac12}}{
\Gamma(\nu-\kappa+1)(p+\frac12a)^{\mu+\nu+\frac12}}
\nnb \\ && \times
{}_2F_1\left(
\mu+\nu+\frac12,\mu-\kappa+\frac12,\nu-\kappa+1;
\frac{p-\frac12a}{p+\frac12a}
\right)\,,
\eea
with Re$[\nu\pm\mu]>-\frac12$ and Re$\left[p+\frac12a\right]>0$,
and we have
\bea
\Gamma_{3(1)}^{(pp)}(p,q) &=& C_\eta
\int_0^\infty dr H^+_0(\eta,pr)j_0(\frac12qr)e^{-\gamma r}
\nnb \\ &=&
C_\eta
e^{\frac{\pi}{2}\eta + i\sigma_0}
(-2ip)\sum_{n=0}^\infty \frac{(-1)^n}{(\gamma -ip)^{2n+2}}
\left(
\frac{q}{2}
\right)^{2n}
\frac{\Gamma(2n+1)}{\Gamma(2n+2+i\eta)}
\nnb \\ && \times
_2F_1\left(2n+2,1+i\eta,2n+2+i\eta;
\frac{\gamma+ip}{\gamma-ip}\right)
\,.
\eea

The three point vertex function $\Gamma_{3(2)}^{(pp)}(p,q)$ 
is represented by using the result of $\Gamma_{3(1)}^{(pp)}(p,q)$ and
an integral,
\bea
I_5 &=& \int_0^\infty dr H_0^+(\eta,pr)j_1(\frac12qr)\left(
\frac{1}{r} + \gamma
\right)e^{-\gamma r}\,,
\eea
and we have
\bea
I_5 &=& -2ip e^{\frac{\pi}{2}\eta+i\sigma_0}
\sum_{n=1}^\infty (-1)^n\left[
\frac{1}{(2n+1)!} -\frac{1}{(2n)!}
\right]\left(
\frac{q}{2}
\right)^{2n-1}
\nnb \\ && \times \left[
\frac{\gamma}{(\gamma-ip)^{2n-1}}
\frac{\Gamma(2n+1)\Gamma(2n)}{\Gamma(2n+1+i\eta)}
{}_2F_1\left(
2n+1,1+i\eta,2n+1+i\eta;
\frac{\gamma+ip}{\gamma-ip}
\right)
\right. \nnb \\ && \left.
+ \frac{1}{(\gamma -ip)^{2n}}
\frac{\Gamma(2n)\Gamma(2n-1)}{\Gamma(2n+i\eta)}
{}_2F_1\left(
2n,1+i\eta,2n+i\eta;
\frac{\gamma+ip}{\gamma-ip}
\right)
\right]\,.
\eea

\section*{Appendix B}

Using the spin summation relations,
\bea
\sum_{spins}\epsilon_{(d)i}\epsilon_{(d)j}^* &=& \delta_{ij}\,,
\\
\sum_{spins}\epsilon_{(l)}^\alpha \epsilon_{(l)}^{*\beta} &=&
8\left(
k'^\alpha k^\beta + k'^\beta k^\alpha - g^{\alpha\beta}k'\cdot k
\pm i \epsilon^{\alpha\beta\mu\nu}k'_\mu k_\nu
\right)\,,
\eea
where the $\pm$ signs correspond to the initial neutrino and
the antineutrino states, respectively, we have
the spin summation relations for $^1$S$_0$ channel as 
\bea
\sum_{spins} \left|
\vec{\epsilon}_{(l)}\cdot \vec{\epsilon}_{(d)}
\right|^2 &=&  8\left(
3E'E - \vec{k}'\cdot\vec{k}
\right)\,,
\\
\sum_{spins}
\vec{\epsilon}_{(l)}^*\cdot\vec{\epsilon}_{(d)}^*
v\cdot\epsilon_{(l)}\hat{q}\cdot\vec{\epsilon}_{(d)}
&=& -8\left[
E'\hat{q}\cdot\vec{k} + E\hat{q}\cdot\vec{k}'
\pm i\hat{q}\cdot\left(
\vec{k}'\times \vec{k}
\right)
\right]\,,
\\
\sum_{spins}
v\cdot\epsilon_{(l)}^*\hat{q}\cdot\vec{\epsilon}_{(d)}^*
\vec{\epsilon}_{(l)}\cdot\vec{\epsilon}_{(d)}
&=& -8\left[
E'\hat{q}\cdot\vec{k} + E\hat{q}\cdot\vec{k}'
\mp i\hat{q}\cdot\left(
\vec{k}'\times \vec{k}
\right)
\right]\,,
\\
-\sum_{spins}
\vec{\epsilon}_{(l)}^*\cdot \vec{\epsilon}_{(d)}^*
i\hat{q}\cdot\left(
\vec{\epsilon}_{(d)}\times\vec{\epsilon}_{(l)}
\right)
&=& \mp16\left(
E'\hat{q}\cdot \vec{k}
-E\hat{q}\cdot\vec{k}'
\right)\,,
\\
\sum_{spins}
i\hat{q}\cdot\left(
\vec{\epsilon}_{(d)}^*\times\vec{\epsilon}_{(l)}^*
\right)
\vec{\epsilon}_{(l)}\cdot \vec{\epsilon}_{(d)}
&=& \mp16\left(
E'\hat{q}\cdot \vec{k}
-E\hat{q}\cdot\vec{k}'
\right)\,,
\\
\sum_{spins}\left|
v\cdot \epsilon_{(l)}\hat{q}\cdot \vec{\epsilon}_{(d)}
\right|^2  &=&
8\left(
E'E+\vec{k}'\cdot\vec{k}
\right)\,,
\\
\sum_{spins} \left|
\hat{q}\cdot\left(
\vec{\epsilon}_{(d)}\times \vec{\epsilon}_{(l)}
\right)
\right|^2 &=&
16\left(
E'E - \hat{q}\cdot\vec{k}'\hat{q}\cdot\vec{k}
\right)\,,
\\
\sum_{spins}
v\cdot\epsilon_{(l)}^*\hat{q}\cdot\vec{\epsilon}_{(d)}^*
i \hat{q}\cdot\left(\vec{\epsilon}_{(d)}\times \vec{\epsilon}_{(l)}\right)
&=& 
\sum_{spins}
i \hat{q}\cdot\left(\vec{\epsilon}_{(d)}^*\times \vec{\epsilon}_{(l)}^*\right)
v\cdot\epsilon_{(l)}\hat{q}\cdot\vec{\epsilon}_{(d)}
= 0\,.
\eea

For the $^3$S$_1$ channel, we have
\bea
\sum_{spins} \left|
v\cdot\epsilon_{(l)}\vec{\epsilon}^*\cdot\vec{\epsilon}_{(d)}
\right|^2 &=& 24\left(
E'E + \vec{k}'\cdot\vec{k}
\right)\,,
\\
\sum_{spins}
v\cdot \epsilon_{(l)}^*\vec{\epsilon}\cdot \vec{\epsilon}_{(d)}^*
\hat{q}\cdot\vec{\epsilon}_{(l)}\vec{\epsilon}^*\cdot \vec{\epsilon}_{(d)}
&=& -24\left[
E'\hat{q}\cdot\vec{k} +E\hat{q}\cdot\vec{k}'
\mp i\hat{q}\cdot\left(
\vec{k}'\times\vec{k}
\right)
\right]\,,
\\
\sum_{spins}
\hat{q}\cdot\vec{\epsilon}_{(l)}^*\vec{\epsilon}\cdot \vec{\epsilon}_{(d)}^*
v\cdot \epsilon_{(l)}\vec{\epsilon}^*\cdot \vec{\epsilon}_{(d)}
&=& -24\left[
E'\hat{q}\cdot\vec{k} +E\hat{q}\cdot\vec{k}'
\pm i\hat{q}\cdot\left(
\vec{k}'\times\vec{k}
\right)
\right]\,,
\\
\sum_{spins}\left|
\hat{q}\cdot \vec{\epsilon}_{(l)}\vec{\epsilon}^*\cdot \vec{\epsilon}_{(d)}
\right|^2 &=& 24\left(
E'E -\vec{k}'\cdot\vec{k} - 2 \hat{q}\cdot\vec{k}'\hat{q}\cdot\vec{k}
\right)\,.
\eea

For the $^3$P$_0$ channel, the terms at LO are the same as those
for the $^1$S$_0$ channel.

For the $^3$P$_1$ channel at LO, using a spin summation relation for the polarization
vector $\epsilon_i$ for the $J=1$ state, 
\bea
\sum_{spins} \epsilon_i\epsilon_j^* = \delta_{ij}\,,
\eea
we have
\bea
&&
\sum_{spins} \left|
iv\cdot \epsilon_{(l)}\vec{\epsilon}^*\cdot\left(
\hat{q}\times \vec{\epsilon}_{(d)}
\right)
\right|^2 = 16\left(
E'E + \vec{k}'\cdot\vec{k}
\right)\,,
\\
&&
\sum_{spins} \left|
\vec{\epsilon}^*\cdot\vec{\epsilon}_{(l)}\hat{q}\cdot\vec{\epsilon}_{(d)}
-\vec{\epsilon}^*\cdot\vec{\epsilon}_{(d)}\hat{q}\cdot\vec{\epsilon}_{(l)}
\right|^2 =  16\left(
2E'E -2\vec{k}'\cdot\vec{k}
-\hat{q}\cdot\vec{k}'\hat{q}\cdot\vec{k}
\right)\,,
\\ 
\lefteqn{
\sum_{spins}
iv\cdot \epsilon_{(l)}\vec{\epsilon}^*\cdot\left(
\hat{q}\times \vec{\epsilon}_{(d)}
\right)\left(
\vec{\epsilon}\cdot\vec{\epsilon}_{(l)}^*\hat{q}\cdot\vec{\epsilon}_{(d)}^*
-\vec{\epsilon}\cdot\vec{\epsilon}_{(d)}^*\hat{q}\cdot\vec{\epsilon}_{(l)}^*
\right) }
\nnb \\ 
\lefteqn{
=\sum_{spins}
iv\cdot \epsilon_{(l)}^*\vec{\epsilon}\cdot\left(
\hat{q}\times \vec{\epsilon}_{(d)}^*
\right)\left(
\vec{\epsilon}^*\cdot\vec{\epsilon}_{(l)}\hat{q}\cdot\vec{\epsilon}_{(d)}
-\vec{\epsilon}^*\cdot\vec{\epsilon}_{(d)}\hat{q}\cdot\vec{\epsilon}_{(l)}
\right) 
= 0\,.
}
\eea

For the $^3$P$_2$ channel at LO, using the spin summation relation for 
the symmetric tensor $\epsilon_{ij}$ for the $J=2$ state, 
\bea
\sum_{spins}\epsilon_{ij}\epsilon_{xy}^* &=& 
\frac12\left(
\delta_{ix}\delta_{jy}
+\delta_{iy}\delta_{jx}
-\frac23\delta_{ij}\delta_{xy}
\right)\,,
\eea
we have
\bea
\sum_{spins}\left|
v\cdot \epsilon_{(l)}\epsilon^{ij*}\hat{q}^i\epsilon_{(d)}^j
\right|^2 &=& \frac{40}{3}\left(
E'E + \vec{k}'\cdot\vec{k}
\right)\,,
\\
\sum_{spins}\left|
i\epsilon^{ij*}\hat{q}^i\epsilon^{jkl}\epsilon_{(l)}^k\epsilon_{(d)}^l
\right|^2 &=& \frac43\left(
20E'E - 6\vec{k}'\cdot\vec{k}
+ 2\hat{q}\cdot\vec{k}'\hat{q}\cdot\vec{k}
\right)\,,
\\
\sum_{spins}
v\cdot \epsilon_{(l)}\epsilon^{ij*}\hat{q}^i\epsilon_{(d)}^j
\epsilon^{ij}\hat{q}^i\epsilon^{jkl}\epsilon_{(l)}^{k*}\epsilon_{(d)}^{l*}
&=&
\sum_{spins}
v\cdot \epsilon_{(l)}^*\epsilon^{ij}\hat{q}^i\epsilon_{(d)}^{j*}
\epsilon^{ij*}\hat{q}^i\epsilon^{jkl}\epsilon_{(l)}^{k}\epsilon_{(d)}^{l}
= 0\,.
\eea

\section*{Appendix C}

Recently, De-Leon, Platter, and Gazit studied the tritium $\beta$ decay 
in pionless EFT and reported two values of $l_{1,A}$ fitted 
to the tritium lifetime as 
$l_{1,A}^{ERE}$ and $l_{1,A}^Z$ in Eqs.~(38a) and (38b) 
in Ref.~\cite{dlpg-prc}.
%
%
We note that a main difference between that work on a three-nucleon system 
for the triton or $^3$He channel and the present work is treatment of 
the effective range terms in the dressed dibaryon 
propagators in Eqs.~(\ref{eq;Ds}), (\ref{eq;Dt}), and (\ref{eq;Ds_pp});
we resum the effective range terms in the dressed dibaryon propagators 
while, because of a singularity- the so called limit cycle appearing
in the three-nucleon system for the triton or $^3$He 
channel~\cite{bhvk-npa00,ab-jpg10} -    
one needs to expand the effective range terms
and introduce a three-body contact interaction at LO.  
For example, the dressed  dibaryon propagator for deuteron channel 
in Eq.~(\ref{eq;Dt}) is expanded as
\bea
D_t(p) &=& \frac{1}{-\gamma -ip+\frac12\rho_d(\gamma^2+p^2)}
= \frac{1}{-\gamma-ip}\left[
1 + \frac12\rho_d (\gamma - ip)
\right] + O(\rho_d^2)\,.
\eea
For a two-nucleon system, e.g., the deuteron, 
one has $p=i\gamma$, and the two terms in the bracket 
in the above equation become $1+\gamma \rho_d\simeq 1+0.4$.  
This is the well-known slow converging effective range correction, 
$\gamma\rho_d\simeq 1/3$, in the deuteron channel. 
For a three-nucleon system, 
one has $p = i\sqrt{-m_N E + \frac14\vec{l}^2-i\epsilon}$, where 
$E$ is the total energy and $\vec{l}$ is the loop momentum
for the three-nucleon system. 
Thus, for the triton, 
one has $1+\frac12\rho_d(\gamma + \gamma_3) \simeq 1+ 0.6$
where $\gamma_3 = \sqrt{m_NB_3+\frac14\vec{l}^2}$; 
$B_3$ is the triton binding energy, $B_3=8.48$~MeV, and 
we assumed $\vec{l}=0$. Thus, one has an expansion parameter
due to the effective range expansion for triton channel 
as $\frac12\rho_d(\gamma+\gamma_3)\simeq 3/5$ (or larger);
the convergence of the effective range
terms for the triton channel would be slower than that for 
the deuteron channel. 
Thus, because of the slow convergence of the effective range terms,
to directly fit a value of $l_{1A}$ to the result from 
the calculation of a three-nucleon system, the triton $\beta$ decay
in pionless EFT may not be well matched to the present work.


\begin{thebibliography}{}

\bibitem{dgh2014}
J.~F. Donogue, E. Golowich, and B.~R. Holstein,
\textit{Dynamics of the Standard Model},
2nd ed. (Cambridge University Press, Cambridge 2014).

\bibitem{sno1}
Q.~R. Ahmad \textit{et al.},
Phys. Rev. Lett. \textbf{87}, 071301 (2001).

\bibitem{sno2}
Q.~R. Ahmad \textit{et al.},
Phys. Rev. Lett. \textbf{89}, 011301 (2002).

\bibitem{sno3}
Q.~R. Ahmad \textit{et al.},
Phys. Rev. Lett. \textbf{89}, 011302 (2002).

\bibitem{sno4}
B. Aharmim \textit{et al.},
Phys. Rev. \textbf{72}, 055502 (2005).

\bibitem{pas1979}
E. Pasierb, H. S. Gurr, J. Lathrop, F. Reines, and H. W. Sobel,
Phys. Rev. Lett. {\bf 43}, 96 (1979).

\bibitem{ver1991}
A. G. Vershinsky, {\it et al.}, 
JETF Lett. {\bf 53}, 513 (1991).

\bibitem{koz2000}
Yu.V. Kozlov, S.V. Khalturtsev, I.N. Machulin, A.V. Martemyanov, V.P. Martemyanov, S.V. Sukhotin, V.G. Tarasenkov, E.V. Turbin, and V.N. Vyrodov , 
Phys. At. Nucl. {\bf 63}, 1016 (2000).

\bibitem{ril1999}
S. P. Riley, Z. D. Greenwood, W. R. Kropp, L. R. Price,
F. Reines, H. W. Sobel, Y. Declais, A. Etenko, and M.
15 Skorokhvatov, Phys. Rev. C {\bf 59}, 1780 (1999).

\bibitem{for2012}
J. A. Formaggio, and G. P. Zeller,
Rev. Mod. Phys. {\bf 84}, 1307 (2012).

\bibitem{ku-prl66}
F.~J. Kelly and H. Uberall,
Phys. Rev. Lett. \textbf{16}, 145 (1966).

\bibitem{eb-npa68}
S.~D. Ellis and J.~N. Bahcall,
Nucl. Phys. A \textbf{114}, 636 (1968).

\bibitem{tkk-prc90}
N. Tatara, Y. Kohyama, and K. Kubodera,
Phys. Rev. C \textbf{42}, 1694 (1990).

\bibitem{snpa}
S. Nakamura, T. Sato, V.~P. Gudkov, and K. Kubodera,
Phys. Rev. C {\bf 63}, 034617 (2001), erratum; 
Phys. Rev. C {\bf 73}, 049904 (2006).

\bibitem{netal-npa}
S. Nakamura, T. Sato , S. Ando, T.S. Park, F. Myhrer, V. P. Gudkov, and K. Kubodera,
Nucl. Phys. A \textbf{707}, 561 (2002).

\bibitem{k-npb97}
D.~B. Kaplan,
Nucl. Phys. B \textbf{494}, 471 (1997).

\bibitem{bs-npa01}
S.~R. Beane and M.~J. Savage,
Nucl. Phys. A \textbf{694}, 511 (2001).

\bibitem{prc2005} 
S. Ando, and C.~H. Hyun,
Phys. Rev. C {\bf 72}, 014008 (2005).

\bibitem{prc2006} 
S. Ando, R.~H. Cyburt, S.~W. Hong, and C.~H. Hyun,
Phys. Rev. C {\bf 74}, 025809 (2006).

\bibitem{prc2007} 
S. Ando, J.~W. Shin, C.~H. Hyun, and S.~W Hong,
Phys. Rev. C {\bf 76}, 064001 (2007).

\bibitem{plb2008} 
S. Ando, J.~W. Shin, C.~H. Hyun, S.~W. Hong, and K. Kubodera,
Phys. Lett B {\bf 668}, 187 (2008).

\bibitem{prc2012} 
S.-I. Ando, and C.~H. Hyun,
Phys. Rev. C {\bf 86}, 024002 (2012).

\bibitem{prc2011} 
S.-I. Ando, Y.-H. Song, C.~H. Hyun, and K. Kubodera,
Phys. Rev. C {\bf 83}, 064002 (2011).

\bibitem{fb2013spin} 
Y.-H. Song, C.~H. Hyun, S.-I. Ando, and K. Kubodera,
Few Body Syst. {\bf 54}, 371 (2013).

\bibitem{prc2017} 
Y.-H. Song, S.-I. Ando, and C.~H. Hyun,
Phys. Rev. C {\bf 96}, 014001 (2017).

\bibitem{mpla2009} 
C.~H. Hyun, J.~W. Shin, and S. Ando,
Mod. Phys. Lett. A {\bf 24}, 827 (2009).

\bibitem{prc2010} 
J.~W. Shin, S. Ando, and C.~H. Hyun,
Phys. Rev. C {\bf 81}, 055501 (2010).

\bibitem{npa2010} 
S.-i. Ando, C.~H. Hyun, and J.~W. Shin,
Nucl. Phys. A {\bf 844}, 165 (2010).

\bibitem{fb2013pv} 
J.~W. Shin, S.-I. Ando, C.~H. Hyun, and S.~W. Hong,
Few Body Syst. {\bf 54}, 359 (2013).

\bibitem{prc2013} 
J.~W. Shin, C.~H. Hyun, S.-I. Ando, and S.~W. Hong,
Phys. Rev. C {\bf 88}, 035501 (2013).

\bibitem{nasu}
S. Nasu, S.~X. Nakamura, K. Sumiyoshi, T. Sato, F. Myhrer, and K. Kubodera,
Astrophys. J. {\bf 801}, 78 (2015).

\bibitem{hybrid}
S. Ando, Y.~H. Song, T.-S. Park, H.~W. Fearing, and K. Kubodera,
Phys. Lett. B {\bf 555}, 49 (2003).

\bibitem{butler}
M. Butler, J.-W. Chen, and X. Kong,
Phys. Rev. C {\bf 63}, 035501 (2001).

\bibitem{baroni}
A. Baroni, and R. Schiavilla,
Phys. Rev. C {\bf 96}, 014002 (2017).

\bibitem{msprl1993}
W.~J. Marciano and A. Sirlin, 
Phys. Rev. Lett. \textbf{71}, 3629 (1993).

\bibitem{aetalplb2004}
S. Ando, H.W. Fearing, V.P. Gudkov, K. Kubodera, F. Myhrer, S. Nakamura, and T. Sato ,
Phys. Lett. B \textbf{595}, 250 (2004).

\bibitem{nnfusion}
S. Ando, and K. Kubodera,
Phys. Lett. B {\bf 633}, 253 (2006).

\bibitem{morita1973}
M. Morita, 
{\it Beta Decay and Muon Capture}, (W. A. Benjamin Inc., 1973) p. 27.

\bibitem{rfhpprc2014}
E. Ryberg, C. Forssen, H.-W. Hammer, and L. Platter,
Phys. Rev. C \textbf{89}, 014325 (2014).

\bibitem{aprc2019}
S.-I. Ando, 
Phys. Rev. C \textbf{100}, 015807 (2019).

\bibitem{petal-prc03}
T.-S. Park, L. E. Marcucci, R. Schiavilla, M. Viviani, A. Kievsky, S. Rosati, K. Kubodera,
D. P. Min, and M. Rho,
Phys. Rev. C \textbf{67}, 055206 (2003).

\bibitem{dlpg-prc}
H. De-Leon, L. Platter, and D. Gazit,
Phys. Rev. C \textbf{100}, 055502 (2019). 

\bibitem{NISTHB}
\textit{NIST Handbook of Mathematical Functions}, edited by F. W. J.
Olver et al. (Cambridge University Press, Cambridge, 2010).

\bibitem{bhvk-npa00}
P.~F. Bedaque, H.-W. Hammer, and U. van Kolck,
Nucl. Phys. A \textbf{676}, 357 (2000).

\bibitem{ab-jpg10}
S. Ando and M.~C. Birse,
J. Phys. G \textbf{37}, 105108 (2010). 

\end{thebibliography}
\end{document}